\documentclass[aps,prd,twocolumn,superscriptaddress]{revtex4}
\usepackage{epsfig,epsf}
\usepackage{amsmath}
\usepackage{amsthm}
\usepackage{amsfonts}
\usepackage{amssymb}
\usepackage{dsfont}
\usepackage{multirow}
\usepackage{appendix}
\usepackage{slashed}
\usepackage[active]{srcltx}
\usepackage{psfrag}

\begin{document}

\title{Light double-gluon hybrid states }
\date{\today}
\author{G.~Daylan Esmer}
\affiliation{Department of Physics, Istanbul University, Vezneciler, 34134 Istanbul, T%
\"{u}rkiye}
\author{B.~Barsbay}
\affiliation{Division of Optometry, School of Medical Services and Techniques,Dogus University, Dudullu-\"{U}mraniye, 34775
Istanbul, T\"{u}rkiye}
\author{K.~Azizi}
\thanks{Corresponding author: kazem.azizi@ut.ac.ir}
\affiliation{Department of Physics, University of Tehran, North Karegar Avenue, Tehran
14395-547, Iran}
\affiliation{Department of Physics, Dogus University, Dudullu-\"{U}mraniye, 34775
Istanbul, T\"{u}rkiye}
\author{H.~Sundu}
\affiliation{Department of Physics Engineering, Istanbul Medeniyet University, 34700
Istanbul, T\"{u}rkiye}
\author{S.~T\"{u}rkmen}
\affiliation{Department of Physics, Istanbul University, Vezneciler, 34134 Istanbul, T%
\"{u}rkiye}

\begin{abstract}
We investigate light hybrid mesons composed of a light quark-antiquark pair and two gluons 
within the framework of QCD sum rules. We focus on states with quantum numbers 
$J^{\mathrm{PC}} = 0^{++}, 0^{+-}, 0^{-+}, 0^{--}$ and $J^{\mathrm{PC}} = 1^{++}, 1^{+-}, 
1^{-+}, 1^{--}$. By employing various interpolating currents constructed from valence light 
quarks and gluon fields, we determine the masses and current couplings of the $\bar{q}GGq$, 
$\bar{q}GGs$, and $\bar{s}GGs$ hybrid configurations. Nonperturbative effects are 
incorporated through quark and gluon condensates up to dimension twelve in the operator 
product expansion, improving the reliability of the numerical predictions. The results 
presented here may provide useful input for future experimental searches for light hybrid 
mesons and can also serve as a basis for studies of their decay properties and interactions 
with other hadronic states.
\end{abstract}

\maketitle


\section{Introduction}

\label{sec:Intro}

In the conventional quark model ~\cite{ParticleDataGroup:2022pth,Gell-Mann:1964ewy,Zweig:1964jf}, hadrons are described as mesons composed of quark–antiquark pairs ($q\bar{q}$) and baryons formed by three quarks ($qqq$). However, quantum chromodynamics (QCD), as the fundamental theory of the strong interaction, predicts a much richer hadron spectrum that extends beyond these conventional configurations. In particular, QCD allows for the existence of exotic hadronic states such as multiquark systems, hybrid mesons, and glueballs, in which the gluonic degrees of freedom play a crucial role in determining their internal dynamics and quantum structure.

A hybrid meson is typically considered to be a bound state containing a quark–antiquark pair and an excited gluonic field ($\bar{q}Gq$). The explicit participation of the gluonic degree of freedom enables these states to exhibit quantum numbers that cannot be realized in ordinary mesons, for instance $J^{\mathrm{PC}}= 0^{--}, 0^{+-}, 1^{-+}, 2^{+-}, 3^{-+}, 4^{+-}, ...$ ~\cite{ParticleDataGroup:2020ssz}. Among the observed light hybrid candidates are $\pi_1(1400)$~\cite{IHEP-Brussels-LosAlamos-AnnecyLAPP:1988iqi}, $\pi_1(1600)$~\cite{E852:2001ikk} and $\pi_1(2015)$~\cite{E852:2004gpn} with the exotic quantum number $J^{PC}= 1^{-+}$. More recently, the isoscalar $\eta_{1}(1855)$ has been reported by the BESIII Collaboration in the decay channel $J/\psi\rightarrow \gamma\eta_{1}\rightarrow \gamma\eta\eta^{\prime}$~\cite{BESIII:2022riz,BESIII:2022iwi}. This state, with $J^{PC}= 1^{-+}$, is regarded as the isoscalar partner of  $\pi_1(1600)$ and a likely member of the light hybrid nonet~\cite{Chen:2022qpd,Chen:2023ukh,Chen:2022isv,Qiu:2022ktc,Shastry:2023ths,Shastry:2022mhk}.

Recent progress in coupled-channel analyses has significantly clarified the long-standing ambiguity surrounding the $\pi_1(1400)$ and $\pi_1(1600)$ resonances, suggesting that existing experimental data can be adequately described by a single $\pi_1(1600)$ state ~\cite{JPAC:2018zyd,Kopf:2020yoa}. Furthermore, lattice QCD studies of radially excited states have identified $\pi_1(2015)$ as a promising candidate for the first excited hybrid meson, providing valuable insight into the structure of gluonic excitations ~\cite{Dudek:2010wm}. The discovery of $\eta_{1}(1855)$ has also opened new perspectives for understanding isoscalar hybrid configurations and the role of nonperturbative gluonic dynamics in hadron formation ~\cite
{Chen:2022qpd,Shastry:2022mhk,Wan:2022xkx,Dong:2022cuw,Yan:2023vbh,Yang:2022rck,Qiu:2022ktc}.

Over the past decades, numerous theoretical investigations have been carried out on the $\bar{q}Gq$ hybrid mesons using a variety of approaches, including flux-tube models~\cite{Isgur:1984bm,Isgur:1985vy,Close:1994hc,Barnes:1995hc,Page:1998gz,Burns:2006wz}, the MIT bag model ~\cite{Barnes:1982tx,Chanowitz:1982qj}, lattice QCD~\cite{Lacock:1996vy,Lacock:1997an,McNeile:1998cp,Lacock:1998be,Bernard:2003jd,Hedditch:2005zf,Dudek:2009qf,Dudek:2010wm,Dudek:2011tt,Dudek:2011bn,Dudek:2013yja} and QCD sum rules~\cite{Balitsky:1982ps,Govaerts:1983ka,Govaerts:1984hc,Govaerts:1985fx,Govaerts:1986pp,Zhu:1998ki,Jin:2002rw,Qiao:2010zh,Berg:2012gd,Chen:2013zia,Chen:2013pya,Chen:2013eha,Li:2021fwk,Palameta:2017ols,Palameta:2018yce,Ho:2018cat,Ho:2019org,Chen:2022qpd,Barsbay:2024vjt,Esmer:2025xss,Agaev:2025llz,Agaev:2025mqe}. A consistent outcome of these studies is the emergence of a characteristic supermultiplet structure. The lowest-lying hybrid configurations involve states with negative parity and positive charge conjugation in the scalar, axial, and tensor channels, together with a vector state carrying negative charge conjugation. In contrast, a higher-lying supermultiplet corresponds to states with positive parity combined with negative charge conjugation, predominantly within the scalar and tensor quantum-number channels~\cite{HadronSpectrum:2012gic,Chen:2013zia,Meyer:2015eta,Dudek:2013yja,Dudek:2009qf,Dudek:2010wm,Dudek:2011tt,Dudek:2011bn,Wang:2023whb}. 

Recently, theoretical attention has extended to hybrid configurations involving two 
valence gluons, often referred to as double-gluon or gluon-rich hybrids. In the light 
quark sector, Su \textit{et al.} (2023) studied these states using QCD sum rules, 
predicting that the mass of the $J^{PC}=2^{+-}$ double-gluon hybrid meson is 
approximately $2.26~\mathrm{GeV}$ for the $\bar{q}GGq$ configuration, while the 
corresponding $\bar{s}GGs$ state has a slightly larger mass of about 
$2.38~\mathrm{GeV}$~\cite{Su:2022fqr}.

Along similar lines, another QCD sum rule study investigated double-gluon hybrid 
states with exotic quantum numbers $J^{PC}=1^{-+}$ and $J^{PC}=3^{-+}$. For the 
$\bar{q}GGq$ configuration, the masses were found to be approximately 
$4.35~\mathrm{GeV}$ for $J^{PC}=1^{-+}$ and $3.02~\mathrm{GeV}$ for $J^{PC}=3^{-+}$, 
while the corresponding $\bar{s}GGs$ hybrid states were predicted to have masses of 
about $4.49~\mathrm{GeV}$ for $J^{PC}=1^{-+}$ and $3.16~\mathrm{GeV}$ for 
$J^{PC}=3^{-+}$~\cite{Su:2023jxb}. These results further support the existence of double-gluon hybrid mesons over a broad mass range and highlight their relevance for exploring exotic QCD dynamics.

In the heavy quark sector, Lian et al. (2024) revisited the masses of double-gluon hybrid mesons with exotic quantum numbers $J^{PC}= 1^{-+}$ and $J^{PC}= 2^{+-}$ in the framework of QCD sum rules. Their analysis predicts that the masses of these states lie in the ranges of  $ 6.1-7.2 $ GeV and $ 6.3-6.4 $ GeV for the charmonium systems, and $ 13.7-14.3 $ GeV and $ 12.6-13.3 $ GeV for the bottomonium systems, respectively~\cite{Lian:2024fsb}. These heavy hybrid mesons are of particular interest for experimental searches at facilities such as Belle II, PANDA, Super-B, GlueX, and LHCb, where they could decay into pairs of charmed or bottom mesons, or into combinations of charmed/bottom and light mesons.

The exploration of light double-gluon hybrid mesons provides an important opportunity 
to advance our understanding of nonperturbative QCD. In the light quark sector, the 
explicit inclusion of dynamical gluon degrees of freedom is expected to play a 
significant role in shaping the properties of hadronic states. Studies of such hybrid 
configurations can therefore offer valuable insight into color confinement, gluon 
self-interactions, and the emergence of exotic hadronic structures beyond the 
conventional quark model. 

This article is structured in the following from: In Sec.~\ref{sec:Hqqbarggmass}, we 
investigate double-gluon hybrid mesons with various $J^{PC}$ quantum numbers within the QCD 
sum rule formalism. By employing various interpolating currents constructed from valence 
light quarks and gluon fields, we determine the masses and current couplings of the $\bar{q}
GGq$, $\bar{q}GGs$, and $\bar{s}GGs$ hybrid configurations. Among the eight interpolating 
currents considered, four are found to vanish identically due to internal symmetry 
properties of the gluon fields. In Sec.~\ref{sec:Numeric}, we perform the numerical analysis 
of the remaining nonvanishing sum rules, with the results summarized and discussed in 
Sec.~\ref{sec:Dis}.


\section{Spectroscopic Parameters}

\label{sec:Hqqbarggmass}

The QCD sum rules for the parameters of the light double-gluon hybrid states
$\bar{q}GGq$, $\bar{q}GGs$, and $\bar{s}GGs$ are obtained from the analysis of the
two-point correlation functions corresponding to the quantum numbers 
$J^{\mathrm{PC}} = 0^{++},0^{+-}, 0^{-+},0^{--}$ and $J^{\mathrm{PC}} = 1^{++},1^{+-}, 1^{-+},1^{--}$. 
For the scalar and pseudoscalar channels, the correlation function is defined as
\begin{equation}
\Pi(q)=i\int d^{4}x\,e^{iqx}\langle 0|\mathcal{T}\{J(x)J^{\dagger}(0)\}|0\rangle ,
\label{eq:CF1}
\end{equation}
whereas for the vector and axial-vector channels we consider
\begin{equation}
\Pi_{\alpha\beta,\alpha^{\prime}\beta^{\prime}}(q)
=i\int d^{4}x\,e^{iqx}
\langle 0|\mathcal{T}\{J_{\alpha\beta}(x)
J_{\alpha^{\prime}\beta^{\prime}}^{\dagger}(0)\}|0\rangle .
\label{eq:CF2}
\end{equation}

Some of the interpolating currents for the double-gluon hybrid states have been constructed in Ref.~\cite{Chen:2021smz}.

For completeness, we list them below:
\begin{eqnarray}
J_{0^{++}} &=& \bar q_a \gamma_5 \lambda_n^{ab} q_b~d^{nrs}~g_s^2 G_r^{\mu\nu} \tilde G_{s,\mu\nu} \, ,
\label{def:0pp}
\\
J_{0^{+-}} &=& \bar q_a \gamma_5 \lambda_n^{ab} q_b~f^{nrs}~g_s^2 G_r^{\mu\nu} \tilde G_{s,\mu\nu} \, ,
\label{def:0pn}
\\
J_{0^{-+}} &=& \bar q_a \gamma_5 \lambda_n^{ab} q_b~d^{nrs}~g_s^2 G_r^{\mu\nu} G_{s,\mu\nu} \, ,
\label{def:0np}
\\
J_{0^{--}} &=& \bar q_a \gamma_5 \lambda_n^{ab} q_b~f^{nrs}~g_s^2 G_r^{\mu\nu} G_{s,\mu\nu} \, ,
\label{def:0nn}
\\ \nonumber
J^{\alpha\beta}_{1^{++}} &=& \bar q_a \gamma_5 \lambda_n^{ab} q_b~d^{nrs}~g_s^2 G_r^{\alpha\mu} \tilde G_{s,\mu}^\beta - \{ \alpha \leftrightarrow \beta \} \, ,
\\
\label{def:1pp}
\\ \nonumber
J^{\alpha\beta}_{1^{+-}} &=& \bar q_a \gamma_5 \lambda_n^{ab} q_b~f^{nrs}~g_s^2 G_r^{\alpha\mu} \tilde G_{s,\mu}^\beta - \{ \alpha \leftrightarrow \beta \} \, ,
\\
\label{def:1pn}
\\ \nonumber
J^{\alpha\beta}_{1^{-+}} &=& \bar q_a \gamma_5 \lambda_n^{ab} q_b~d^{nrs}~g_s^2 G_r^{\alpha\mu} G_{s,\mu}^\beta - \{ \alpha \leftrightarrow \beta \} \, ,
\\
\label{def:1np}
\\ \nonumber
J^{\alpha\beta}_{1^{--}} &=& \bar q_a \gamma_5 \lambda_n^{ab} q_b~f^{nrs}~g_s^2 G_r^{\alpha\mu} G_{s,\mu}^\beta - \{ \alpha \leftrightarrow \beta \} \, .
\\
\label{def:1nn}
\end{eqnarray}
Here $G^n_{\mu\nu}$ and $\tilde G^n_{\mu\nu} = G^{n,\rho\sigma} \times 
\epsilon_{\mu\nu\rho\sigma}/2$ denote the gluon field strength tensor and its dual, $q_a$ ($
\bar q_a$) represents the light quark (antiquark) field, $a=1,\ldots,3$ and $n=1,\ldots,8$ 
are color indices, $\mu,\nu,\rho,\sigma$ are Lorentz indices, and $d^{npq}$ ($f^{npq}$) are 
the totally symmetric (antisymmetric) $\mathrm{SU}(3)$ structure constants, which satisfy the contraction identities
\begin{equation}
d_{abc}\, d_{abd} = \frac{N^2-4}{N}\,\delta_{cd},\ f_{abc}\, f_{abd} = N\,\delta_{cd},  \label{eq:sym&antisym}
\end{equation}
with $N=3$ for $\mathrm{SU}(3)$.

To extract the spectroscopic parameters of the double-gluon hybrid states, the 
corresponding hadronic quantities must be related to the fundamental QCD parameters, such as 
the quark masses, quark-gluon condensates of various nonperturbative mass dimensions, the 
strong coupling constant, and several auxiliary parameters introduced at different stages of 
the calculation, following the standard prescriptions of the QCD sum rule method.
For this purpose, the relevant two-point correlation function is evaluated both in the 
hadronic and in the QCD representations. The QCD sum rules of interest are then obtained by 
identifying the coefficients associated with the independent Lorentz structures in these 
representations. Technically, the hadronic representation in the time-like region is 
constructed by introducing a complete set of hadronic states, carrying the same quantum 
numbers as the interpolating current, between the creation and annihilation currents defined 
in coordinate space. As an illustrative example, we consider the interpolating current $ 
J^{\alpha\beta}_{1^{--}} $, which couples to the vector double-gluon hybrid state $\bar{q}
GGq$ with quantum numbers $J^{\mathrm{PC}} = 1^{--} $. After performing the four-dimensional 
integration over the coordinate $ x $, the hadronic representation of the correlation 
function can be written as
\begin{eqnarray}
&&\Pi _{\alpha\beta,\alpha^{\prime}\beta^{\prime}}^{\mathrm{Phys}}(q)=\frac{\langle 0|J^{\alpha\beta}|H_{V}(q,\varepsilon )\rangle
\langle H_{V}(q,\varepsilon )|J_{\alpha^{\prime}\beta^{\prime}}^{\dagger }|0\rangle }{m_{H_{V}}^{2}-q^{2}}+\ldots,  \notag \\
&&  \label{eq:PhysSide}
\end{eqnarray}%
where $\Pi _{\alpha\beta,\alpha^{\prime}\beta^{\prime}}^{\mathrm{Phys}}(q) $ denotes the physical (hadronic) representation of the correlation function, and $m_{H_{V}}$ is the mass of the light vector double-gluon hybrid state. To proceed, we introduce the following matrix element:
\begin{eqnarray}
\langle 0|J^{\alpha\beta}|H_{V}(q,\varepsilon )\rangle &=& i f_{H_{V}} \varepsilon^{\alpha\beta\mu\nu} \varepsilon_\mu q_\nu ,
\nonumber \\ \langle H_{V}(q,\varepsilon )|J_{\alpha^{\prime}\beta^{\prime}}^{\dagger }|0\rangle  &=& i f_{H_{V}} \varepsilon^{\alpha^{\prime}\beta^{\prime}\mu^{\prime}\nu^{\prime}} \varepsilon^*_{\mu^{\prime}} q_{\nu^{\prime}} .
\label{eq:Mel1}
\end{eqnarray}
Substituting this matrix element and its conjugation into Eq.~\eqref{eq:PhysSide} and summing over the polarization states of the vector hybrid meson,
\begin{equation}
\sum_{\lambda}\varepsilon^{(\lambda)}_\mu \varepsilon^{*(\lambda)}_{\mu^{\prime}}
= - g_{\mu\mu^{\prime}} + \frac{q_\mu q_{\mu^{\prime}}}{m_{H_V}^2},
\end{equation}
we obtain
\begin{eqnarray}
 &&\Pi^{\mathrm{Phys}}_{\alpha\beta,\alpha^{\prime}\beta^{\prime}}(q)
=
-\frac{f_{H_V}^2}{m_{H_V}^2-q^2}\Bigg[-q_\beta q_{\beta^{\prime}}g_{\alpha\alpha^{\prime}} \notag \\
&&+q_\beta q_{\alpha^{\prime}}g_{\alpha\beta^{\prime}}+q_\alpha q_{\beta^{\prime}}g_{\beta\alpha^{\prime}}-q_\alpha q_{\alpha^{\prime}}g_{\beta\beta^{\prime}}\notag \\
&&+q^2(g_{\alpha\alpha^{\prime}}g_{\beta\beta^{\prime}}-g_{\alpha\beta^{\prime}}g_{\beta\alpha^{\prime}})\Bigg]
+\cdots .
\label{eq:PhysExpanded}
\end{eqnarray}

The hadronic representation in Eq.~\eqref{eq:PhysExpanded} is expressed in terms of several independent Lorentz structures. In order to isolate the contribution of the vector double-gluon hybrid state with quantum numbers $J^{\mathrm{PC}} = 1^{--}$ and to ensure a stable matching with the QCD representation, we select the Lorentz structure proportional to $ g_{\alpha\alpha^{\prime}}g_{\beta\beta^{\prime}} $.

On the QCD side, the same correlation function is evaluated by employing
the light-quark propagators within the framework of the operator product
expansion (OPE). In this approach, the time-ordered product of the
interpolating currents is expanded in terms of local operators with
increasing mass dimensions, while their vacuum expectation values encode
the nonperturbative dynamics of QCD through quark and gluon condensates.
Accordingly, the QCD representation of the correlation function takes the form
\begin{eqnarray}
&&\Pi _{\alpha\beta,\alpha^{\prime}\beta^{\prime}}^{\mathrm{OPE}}(q)=i\int
d^{4}xe^{iqx}{\lambda }_{n}^{ab}{\lambda }_{n^{\prime }}^{a^{\prime }b^{\prime }}f^{nrs}f^{n^{\prime }r^{\prime }s^{\prime }}g_{s}^{4} \notag \\
&&\times \langle 0|(G_r^{\alpha\mu}(x)G_{s,\mu}^\beta(x)-G_r^{\beta\mu}(x)G_{s,\mu}^\alpha(x)) \notag \\
&&\times(G_{s^{\prime },\mu^{\prime }}^{\beta^{\prime }}(0)G_{r^{\prime }}^{\alpha^{\prime }\mu^{\prime }}(0)-G_{s^{\prime },\mu^{\prime }}^{\alpha^{\prime }}(0)G_{r^{\prime }}^{\beta^{\prime }\mu^{\prime }}(0))|0\rangle \notag \\
&&\times \mathrm{Tr}\left[ \gamma _{5}S_{q}^{bb^{\prime }}(x)\gamma
_{5}S_{q}^{a^{\prime }a}(-x)\right] ,  \label{eq:QCD}
\end{eqnarray}
where $t_n={\lambda_n}/2$. The explicit form of the light-quark propagator $S_q(x)$ is given in Ref.~\cite{Barsbay:2024vjt}. The function $ \Pi _{\alpha\beta,\alpha^{\prime}\beta^{\prime}}^{\mathrm{OPE}}(q) $ constitutes the QCD side of the sum rules.

The QCD side of the correlation function
$\Pi_{\alpha\beta,\alpha^{\prime}\beta^{\prime}}^{\mathrm{OPE}}(q)$
is constructed by evaluating the operator product expansion in the deep Euclidean region. Two types of contributions are taken into account. The first one arises from tracing over the light-quark fields. In this case, the light-quark propagators generate perturbative terms together with nonperturbative contributions induced by background gluon fields. After performing the vacuum expectation values, these terms lead to gluon condensate contributions, which are implemented through the local matrix element of two gluon field-strength tensors,
\begin{eqnarray}
\label{eq:Gcond}
&&\langle 0 |G_r^{\alpha\mu}(x)G_{s^{\prime },\mu^{\prime }}^{\beta^{\prime }}(0)|0 \rangle =
\frac{\langle G^2\rangle }{96}\delta^{rs^{\prime }} [g_{\alpha \beta^{\prime }} g_{\mu \mu^{\prime }} \notag \\
&&-g_{\alpha \mu^{\prime }} g_{\beta^{\prime }\mu }].
\end{eqnarray}
The second contribution originates from gluonic matrix elements at finite separation and is treated by employing the full coordinate-space gluon propagator,
\begin{eqnarray}
\label{eq:Gprop}
&&\langle 0 |G_r^{\alpha\mu}(x)G_{s^{\prime },\mu^{\prime }}^{\beta^{\prime }}(0)|0 \rangle  =
\frac{\delta^{rs^{\prime }}}{2 \pi^2 x^4} [g_{\mu \mu'}(g_{\alpha \beta^{\prime }}-\frac{4 x_{\alpha} x_{\beta^{\prime }}}{x^2}) \nonumber \\
&& +(\mu, \mu^{\prime }) \leftrightarrow (\alpha, \beta^{\prime })
 -\mu \leftrightarrow \alpha -\mu^{\prime } \leftrightarrow \beta^{\prime }].
\end{eqnarray}
Both contributions are consistently incorporated into the OPE, and their combined effects are taken into account in the construction of the QCD sum rules.

The correlator $ \Pi _{\alpha\beta,\alpha^{\prime}\beta^{\prime}}^{\mathrm{OPE}}(q) $  involves several Lorentz structures discussed above. In order to derive the QCD sum 
rules for the mass $ m_{H_{V}} $ and the current coupling $f_{H_{V}}$, we isolate the 
invariant amplitude $\Pi ^{\mathrm{OPE}}(q^{2})$ associated with the term $ 
g_{\alpha\alpha^{\prime}}g_{\beta\beta^{\prime}} $. By equating $\Pi ^{\mathrm{OPE}}(q^{2})$ 
with $\Pi ^{\mathrm{Phys}}(q^{2})$ and subsequently applying the Borel transformation and 
continuum subtraction, we obtain
\begin{equation}
m_{H_{V}}^{2}=\frac{\Pi ^{\prime }(M^{2},s_{0})}{\Pi (M^{2},s_{0})},  \label{eq:Mass}
\end{equation}
and
\begin{equation}
f_{H_{V}}^{2}=\frac{e^{m_{H_{V}}^{2}/M^{2}}}{m_{H_{V}}^{2}}%
\Pi (M^{2},s_{0}),  \label{eq:Coupl}
\end{equation}
where $\Pi ^{\prime }(M^{2},s_{0})=d\Pi (M^{2},s_{0})/d(-1/M^{2})$. Above $
\Pi (M^{2},s_{0})$ is the amplitude $\Pi ^{\mathrm{OPE}}(p^{2})$ obtained
after the Borel transformation and continuum subtraction procedures. It is a
function of the Borel and continuum subtraction parameters $M^{2}$ and $
s_{0} $. The equalities Eqs.\ (\ref{eq:Mass}) and (\ref{eq:Coupl}) are the
sum rules for the mass and current coupling of the vector hybrid meson. In the present work, we calculate $\Pi (M^{2},s_{0})$ by taking into account quark, gluon and mixed vacuum condensates up to dimension 12, and the explicit form of this correlator is presented in the Appendix. It has the following form
\begin{equation}
\Pi (M^{2},s_{0})=\int_{4m_{q}^{2}}^{s_0}ds\rho ^{
\mathrm{OPE}}(s)e^{-s/M^2}+\Pi (M^{2}),  \label{eq:CF3}
\end{equation}
where $\rho ^{\mathrm{OPE}}(s)$ is the two-point spectral density. The second component of 
the invariant amplitude $\Pi (M^{2})$ contains nonperturbative contributions calculated 
directly from the OPE side of each case. Owing to internal symmetry properties of the gluon 
fields entering the corresponding correlation functions, the interpolating currents $ 
J_{0^{--}}, J_{0^{+-}}, J_{1^{-+}}$ and $ J_{1^{++}} $ are found to vanish identically 
within the present framework. The same methodology is systematically applied to the 
remaining interpolating currents in order to extract the QCD sum rules for the physical 
properties of the corresponding hybrid states.

\section{Numerical Results}
\label{sec:Numeric}

This section presents a numerical analysis of the masses and current couplings, using quark masses from the PDG and the quark-gluon condensates as follows:
\begin{eqnarray}
&&\ m_{u}=(2.16\pm 0.04)~\mathrm{MeV},\
\notag \\
&&\ m_{d}=(4.70\pm 0.04)~\mathrm{MeV},\
\notag \\
&&\ m_{s}=(93.5\pm 0.5)~\mathrm{MeV},\
\notag \\
&&\langle \overline{q}q\rangle =-(0.24\pm 0.01)^{3}~\mathrm{GeV}^{3},\
\notag \\
&&\langle \overline{s}s\rangle =0.8~\langle \overline{q}q\rangle ,\
\notag \\
&&\langle \overline{q}g_{s}\sigma Gq\rangle =m_{0}^{2}\langle \overline{q}
q\rangle ,\ m_{0}^{2}=(0.8\pm 0.1)~\mathrm{GeV}^{2},\   \notag \\
&&\langle \frac{\alpha _{s}G^{2}}{\pi }\rangle =(0.012\pm 0.004)~\mathrm{GeV}
^{4}.
 \label{eq:Parameters}
\end{eqnarray}

The QCD sum rule predictions exhibit an explicit dependence on the Borel parameter $M^{2}$ 
and the continuum threshold $s_{0}$. The values of $M^{2}$ and $s_{0}$ appearing in Eqs.\ 
(\ref{eq:Mass}) and (\ref{eq:Coupl}) vary from one process to another and must be fixed in 
accordance with the standard criteria adopted in sum rule analyses. In particular, an 
appropriate choice of these parameters is required to guarantee the dominance of the ground-
state pole contribution ($\mathrm{PC}$) in the correlation functions. Furthermore, the 
convergence of the OPE, together with the stability of the 
extracted observables under variations of the Borel parameter $M^{2}$, constitutes essential 
conditions. To assess the fulfillment of these constraints, we define the pole contribution 
as
\begin{equation}
\mathrm{PC}=\frac{\Pi (M^{2},s_{0})}{\Pi (M^{2},\infty )},  \label{eq:PC}
\end{equation}
and introduce the ratio
\begin{equation}
R(M^{2})=\frac{\Pi ^{\mathrm{DimN}}(M^{2},s_{0})}{\Pi (M^{2},s_{0})},
\label{eq:Conv}
\end{equation}
where $\Pi^{\mathrm{DimN}}(M^{2},s_{0})= \sum_{N=8}^{12}\Pi^{\mathrm{DimN}}$
denotes the contribution of the highest-dimensional operators retained in the OPE.
These terms arise solely from light-quark propagators and local two-gluon operators,
and lead to higher-dimensional four-quark and gluonic condensates.

\begin{figure}[h]
\includegraphics[width=8.5cm]{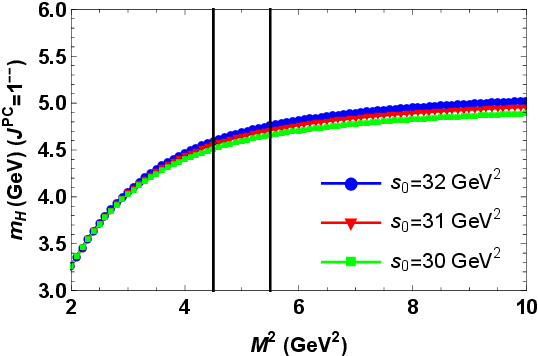}
\caption{The mass $m_{H}$ as a function of the Borel parameter $M^{2}$ for different values of $s_{0}$. The two vertical lines indicate the boundaries of the Borel window,
within which the constraints imposed on $\Pi(M^{2},s_{0})$ are satisfied.}
\label{fig:WR}
\end{figure}

\begin{figure}[h]
\includegraphics[width=8.5cm]{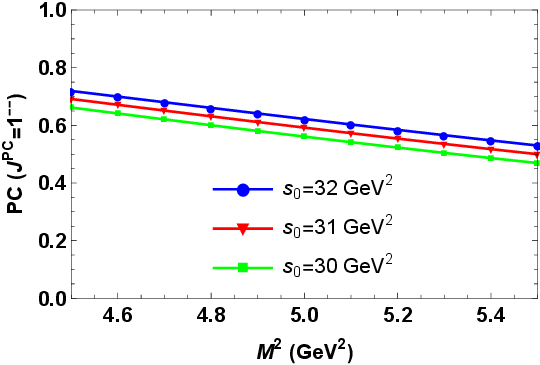}
\caption{Pole contribution $\mathrm{PC}$ as a function of $M^{2}$ at fixed $s_{0}$.}
\label{fig:PC}
\end{figure}

We now focus on the vector $\bar{q}GGq$ hybrid state. Our numerical analysis indicates that, for the vector state $H_{V}$, the working regions of the Borel parameter $M^{2}$ and the continuum threshold $s_{0}$, as shown in Fig.~\ref{fig:WR}, are determined as
\begin{equation}
M^{2}\in \lbrack 4.5,5.5]~\mathrm{GeV}^{2},\ s_{0}\in \lbrack 30,32]~\mathrm{
GeV}^{2},  \label{eq:Wind1}
\end{equation}
which satisfy the standard constraints of the QCD sum rule approach. Within these intervals, the pole contribution takes the values $\mathrm{PC}\simeq 0.50$ and $\mathrm{PC}\simeq 0.69$ at $M^{2}=5.5~\mathrm{GeV}^{2}$ and $M^{2}=4.5~\mathrm{GeV}^{2}$, respectively, averaged over the chosen range of $s_{0}$. At the lower bound $M^{2}=4.5~\mathrm{GeV}^{2}$, the nonperturbative contribution is negative and amounts to less than $1\%$ of the total result. 
The dependence of the pole contribution on the Borel parameter $M^{2}$ is 
displayed in Fig.\ \ref{fig:PC}. As a result, for the mass $ m_{H_{V}} $ and the current coupling $f_{H_{V}}$ of the vector state $H_{V}$, we get
\begin{eqnarray}
m_{H_{V}} &=&(4.64\pm 0.13)~\mathrm{GeV},\ \   \notag \\
f_{H_{V}} &=&(0.08\pm 0.01)~\mathrm{GeV}^{5}.
\end{eqnarray}
The dependence of the mass $m_{H}$ and current coupling $f_{H}$ on the Borel 
parameter and continuum threshold is depicted in Figs.\ \ref{fig:Mass} and 
\ref{fig:Decay constant}.

Having presented the detailed numerical results for the vector $\bar{q}GGq$ 
hybrid state, we now extend the analysis to the full set of 
$\bar{q}GGq$, $\bar{q}GGs$, and $\bar{s}GGs$ hybrid configurations. 
For these systems, we consider states with quantum numbers 
$J^{\mathrm{PC}} = 0^{++}, 0^{+-}, 0^{-+}, 0^{--}$ and 
$J^{\mathrm{PC}} = 1^{++}, 1^{+-}, 1^{-+}, 1^{--}$. 
The corresponding working windows of the Borel parameter $M^{2}$ and continuum 
threshold $s_{0}$, together with the pole contributions, convergence ratios 
$R(M^{2})$, extracted masses and current couplings, as well as available 
results from the literature, are collected in Table~\ref{tab:HGG}.

\begin{widetext}

\begin{figure}[h!]
\begin{center}
\includegraphics[totalheight=6cm,width=8cm]{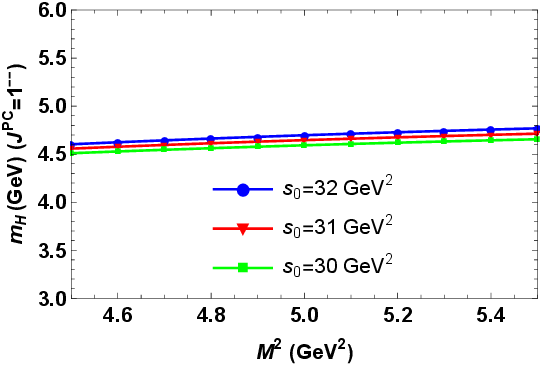}\,\, 
\includegraphics[totalheight=6cm,width=8cm]{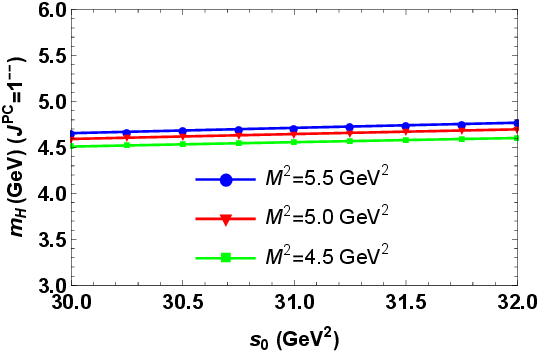}
\end{center}
\caption{ The mass of the $\bar{q}GGq$ hybrid meson as functions of  $M^2$  and  $s_0$.}
\label{fig:Mass}
\end{figure}

\begin{figure}[h!]
\begin{center}
\includegraphics[totalheight=6cm,width=8cm]{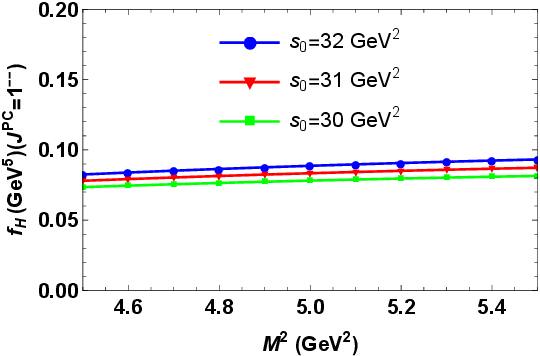}\,\, 
\includegraphics[totalheight=6cm,width=8cm]{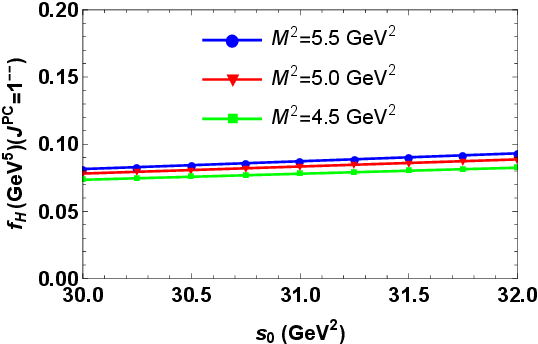}
\end{center}
\caption{ The current coupling of  the $\bar{q}GGq$ hybrid meson as functions of  $M^2$  and  $s_0$.}
\label{fig:Decay constant}
\end{figure}

\begin{table}[tbp]
\begin{tabular}{|c|c|c|c|c|c|c|c|c|}
\hline\hline
$J^{\mathrm{PC}}$ & Hybrid & $M^2~(\mathrm{GeV}^2)$ & $s_0~(\mathrm{GeV}^2)$ & $%
\mathrm{PC} (\%)$ &$ \left| R(M^{2}) \right| (\%) $& Mass $(\mathrm{GeV})$ & $f (\mathrm{GeV}^5)$
& Mass $(\mathrm{GeV})$ \cite{Su:2022fqr} \\ \hline
$0^{++}$ & $\bar{q}GGq$ & $4.5-5.5$ & $30-32$ & $69-50$ & $<1.6$ & $4.63\pm 0.13$&$0.83\pm 0.12$&$ 5.61^{+0.29}_{-0.27} $ \\
$0^{-+}$ & $\bar{q}GGq$ & $4.5-5.5$ & $30-32$ & $69-50$ & $<1.5$ & $4.63\pm 0.13$&$0.83\pm 0.12$&$ 4.25^{+0.32}_{-0.39} $ \\
$1^{+-}$ & $\bar{q}GGq$ & $4.5-5.5$ & $30-32$ & $69-50$ & $<0.7$ & $4.64\pm 0.13$&$0.08\pm 0.01$&$ 5.46^{+0.25}_{-0.18} $  \\ 
$1^{--}$ & $\bar{q}GGq$ & $4.5-5.5$ & $30-32$ & $69-50$ & $<0.6$ & $4.64\pm 0.13$&$0.08\pm 0.01$&$ 3.74^{+0.30}_{-0.35} $  \\ \hline
$0^{++}$ & $\bar{q}GGs$ & $4.7-5.7$ & $31-33$ & $66-50$ & $<1.1$ & $4.77\pm 0.11$&$0.93\pm 0.12$&$ - $ \\
$0^{-+}$ & $\bar{q}GGs$ & $4.7-5.7$ & $31-33$ & $68-50$ & $<1.1$ & $4.72\pm 0.12$&$0.92\pm 0.13$&$ - $ \\
$1^{+-}$ & $\bar{q}GGs$ & $4.7-5.7$ & $31-33$ & $68-50$ & $<0.4$ & $4.72\pm 0.13$&$0.09\pm 0.01$&$ - $  \\ 
$1^{--}$ & $\bar{q}GGs$ & $4.7-5.7$ & $31-33$ & $68-50$ & $<0.5$ & $4.72\pm 0.12$&$0.09\pm 0.01$&$ -$  \\ \hline

$0^{++}$ & $\bar{s}GGs$& $4.9-5.9$ & $32-34$ & $67-50$ & $<0.7$ & $4.79\pm 0.12$&$1.02\pm 0.14$&$ 5.72^{+0.29}_{-0.32} $ \\
$0^{-+}$ & $\bar{s}GGs$ & $4.9-5.9$ & $32-34$ & $67-50$ & $<0.8$ & $4.79\pm 0.12$&$1.02\pm 0.14$&$ 4.34^{+0.36}_{-0.46} $ \\
$1^{+-}$ & $\bar{s}GGs$ & $4.9-5.9$ & $32-34$ & $67-50$ & $<0.3$ & $4.80\pm 0.12$&$0.09\pm 0.01$&$ 5.52^{+0.29}_{-0.27} $  \\ 
$1^{--}$ & $\bar{s}GGs$& $4.9-5.9$ & $32-34$ & $67-50$ & $<0.4$ & $4.80\pm 0.12$&$0.09\pm 0.01$&$ 3.84^{+0.35}_{-0.44} $\\ \hline\hline
\end{tabular}
\centering
\caption{QCD sum rule predictions for the $\bar{q}GGq$, $\bar{q}GGs$, and $\bar{s}GGs$ 
hybrid states with various $J^{\mathrm{PC}}$ quantum numbers. 
The working windows in $M^{2}$ and $s_{0}$, pole contributions, convergence 
ratios, extracted masses and current couplings, and available literature values 
are also shown.}
\label{tab:HGG}
\end{table}

\end{widetext}

\section{Conclusions}

\label{sec:Dis} 
In this work, we have investigated light double-gluon hybrid mesons with quark-gluon
contents $\bar{q}GGq$, $\bar{q}GGs$, and $\bar{s}GGs$ within the framework of the QCD
sum rule approach. The analysis has been carried out for states with quantum numbers
$J^{\mathrm{PC}} = 0^{++}, 0^{+-}, 0^{-+}, 0^{--}$ and $J^{\mathrm{PC}} = 1^{++}, 1^{+-}, 
1^{-+}, 1^{--}$ by employing interpolating currents constructed from light quark fields and 
two gluons in a color-octet configuration.

Among the eight interpolating currents considered in this study, owing to the internal symmetry properties of the gluon fields entering the corresponding correlation functions, the currents $J_{0^{--}}$, $J_{0^{+-}}$, $J_{1^{-+}}$, 
and $J_{1^{++}}$ vanish identically and therefore do not contribute to the sum rules. Consequently, only the remaining currents give rise to nonvanishing correlation functions and physically meaningful predictions.

For the remaining channels, reliable QCD sum rules are derived by carefully selecting
the Borel mass $M^{2}$ and the continuum threshold $s_{0}$ such that the standard
criteria of pole dominance, OPE convergence, and stability of
the extracted observables are satisfied. The numerical results for the working windows,
pole contributions, convergence ratios, masses, and current couplings are summarized
in Table~\ref{tab:HGG}.

Our analysis indicates that the predicted masses of the double-gluon hybrid states lie
in the range of approximately $4.6$-$4.8~\mathrm{GeV}$, with a systematic increase as
the strange-quark content is introduced. The corresponding current couplings are also
found to be stable within the estimated uncertainties. These results suggest a clear
mass hierarchy among the $\bar{q}GGq$, $\bar{q}GGs$, and $\bar{s}GGs$ configurations,
reflecting the underlying quark-mass effects in the hybrid systems.

A detailed comparison with existing QCD sum rule studies reveals both agreements and significant differences. Previous analyses reported masses of approximately $5.61$~GeV ($0^{++}$), $4.25$~GeV ($0^{-+}$), $5.46$~GeV ($1^{+-}$) , and $3.74$~GeV ($1^{--}$) for the $\bar{q}GGq$ system, and $5.72$~GeV ($0^{++}$), $4.34$~GeV ($0^{-+}$), $5.52$~GeV ($1^{+-}$), and $3.84$~GeV ($1^{--}$) for the $\bar{s}GGs$ configuration~\cite{Su:2022fqr}. As shown explicitly in Table~I, our predicted masses are of comparable magnitude but exhibit noticeable shifts, which can be attributed to differences in the interpolating currents, operator bases, and truncation schemes of the operator product expansion.

It is worth emphasizing that, in the aforementioned works, neither the $\bar{q}GGs$ hybrid states corresponding to the interpolating currents considered in the present study nor the associated current couplings were investigated. In contrast, our analysis provides, for the first time within this current structure, mass predictions and current couplings for the $\bar{q}GGs$ hybrid configurations, thereby extending the scope of previous QCD sum rule studies.

The findings presented here provide quantitative predictions that may serve as useful
input for future experimental searches for light double-gluon hybrid mesons. Moreover,
the extracted masses and couplings can be employed in subsequent studies of decay
properties and interaction mechanisms of hybrid states with conventional hadrons,
thereby contributing to a deeper understanding of nonperturbative QCD dynamics.

\section*{ACKNOWLEDGMENTS}

G.~Daylan~Esmer, H.~Sundu and S.~T\"{u}rkmen gratefully acknowledge the financial support from the Scientific and Technological Research Council of T\"{u}rkiye (TUBITAK) under Grant
No.~125F556. 

\appendix*

\begin{widetext}

\section{ Explicit expressions on the OPE side of the sum rules for the $\bar q GG q$ hybrid state with quantum numbers $J^{PC}=1^{--}$ up to dimension 12}

\renewcommand{\theequation}{\Alph{section}.\arabic{equation}} \label{sec:App}

In this Appendix, we list the explicit forms of the invariant functions $\rho 
^{\mathrm{OPE}} (s)  $ and $ \Pi (M^{2}) $ obtained for the $\bar q GG q$ hybrid state with 
quantum numbers $J^{PC}=1^{--}$. Their expressions are given as:
\begin{equation}
\rho ^{\mathrm{OPE}}(s)=\rho ^{\mathrm{pert.}}(s)+\sum_{N=3}^{10}\rho ^{
\mathrm{DimN}}(s),\ \ \Pi (M^{2})=\sum_{N=10}^{12}\Pi ^{\mathrm{DimN}}(M^{2}).
\label{eq:A1}
\end{equation}

The spectral density is obtained by separating its perturbative and nonperturbative contributions as follows:
\begin{eqnarray}
&&\rho ^{\mathrm{pert.}}(s)=-\frac{g_{s}^{4}s^{4}(28m_{q}^{2}+s)}{2^{12}\cdot 3^{2}\cdot 5\cdot 7\pi
^{6}} ,
\end{eqnarray}
\begin{eqnarray}
&&\rho ^{\mathrm{Dim3}}(s)=\frac{g_{s}^{4}m_{q}\langle \overline{q}q\rangle s^{3}}{2^{7}\cdot 3\cdot 5 \pi
^{4}} ,
\end{eqnarray}
\begin{eqnarray}
&&\rho ^{\mathrm{Dim4}}(s)=\langle \alpha _{s}G^{2}/\pi
\rangle\frac{g_{s}^{4}s^{3}}{2^{13}\cdot 3 \cdot 5\pi ^{4}} ,
\end{eqnarray}
\begin{eqnarray}
&&\rho ^{\mathrm{Dim5}}(s )=-\frac{g_{s}^{4}m_{o}^{2}m_{q}\langle \overline{q}q\rangle s^{2}}{2^{6}\cdot 3^{2}\pi ^{4}} ,
\end{eqnarray}
\begin{eqnarray}
&&\rho ^{\mathrm{Dim6}}(s)=-\frac{g_s^{4}\langle \overline{q}q\rangle ^{2} s}{2^{5}\cdot3^{5}\pi^{4}}
\left[
2\cdot3^{3}\pi^{2}(m_q^{2}+s)
+g_s^{2}s
\right] ,
\end{eqnarray}
\begin{eqnarray}
&&\rho^{\mathrm{Dim7}}(s)
=\langle \alpha _{s}G^{2}/\pi
\rangle \frac{g_s^{4} m_q\langle \bar qq\rangle}{2^{5}\cdot3^{3}\pi^{2}}
s,
\end{eqnarray}
\begin{eqnarray}
&&\rho^{\mathrm{Dim8}}(s)
=\frac{g_s^{4}}{2^{8}\cdot3^{2}\pi^{2}}
\left[
2^{5}\,\langle \bar qq\rangle^{2} m_{o}^{2}s
-3\langle g_s^{2}G^{2}\rangle^{2}(2m_q^{2}+s)
\right],
\end{eqnarray}
and
\begin{eqnarray}
&&\rho^{\mathrm{Dim10}}(s)
=-\frac{g_s^{4} m_{o}^{4}\langle \bar qq\rangle^{2}}{2^{5}\cdot 3\pi^{2}} .
\end{eqnarray}

Components of the function $\Pi (M^{2})$ are:
\begin{eqnarray}
&&\Pi ^{\mathrm{Dim10}}(M^{2})=-\frac{g_s^{4} m_{o}^{4}m_q^{2}\langle \bar qq\rangle^{2}}{2^{5}\cdot 3^{3}\pi^{2}} ,
\end{eqnarray}

\begin{eqnarray}
&&\Pi^{\mathrm{Dim11}}(M^{2})
=-\frac{m_q\langle \bar qq\rangle}{2^{5}\cdot3^{6}\pi^{2}}
\left[
2^{3}g_s^{6}m_{o}^{2}\langle \bar qq\rangle^{2}
+3^{4}\pi^{2}g_s^{4}\langle g_s^{2}G^{2}\rangle^{2}
\right],
\end{eqnarray}

\begin{eqnarray}
&&\Pi^{\mathrm{Dim12}}(M^{2})
=\frac{g_s^{4}}{2^{11}\cdot3^{9}\pi^{2}}
\left[
2^{11} g_s^{4}\langle \bar qq\rangle^{4}
-3^{7}\pi^{2}
\langle \alpha _{s}G^{2}/\pi
\rangle \langle g_s^{2}G^{2}\rangle^{2}
\right].
\end{eqnarray}

\end{widetext}

\renewcommand{\theequation}{\Alph{section}.\arabic{equation}} \label{sec:App}



\begin{thebibliography}{99}


\bibitem{ParticleDataGroup:2022pth}
R.~L.~Workman \textit{et al.} [Particle Data Group],
``Review of Particle Physics,''
PTEP \textbf{2022}, 083C01 (2022).


\bibitem{Gell-Mann:1964ewy}
M.~Gell-Mann,
``A Schematic Model of Baryons and Mesons,''
Phys. Lett. \textbf{8}, 214-215 (1964).


\bibitem{Zweig:1964jf}
G.~Zweig,
``An SU(3) model for strong interaction symmetry and its breaking. Version 2.''

\cite{ParticleDataGroup:2020ssz}

\bibitem{ParticleDataGroup:2020ssz}
P.~A.~Zyla \textit{et al.} [Particle Data Group],
``Review of Particle Physics,''
PTEP \textbf{2020}, no.8, 083C01 (2020).


\bibitem{IHEP-Brussels-LosAlamos-AnnecyLAPP:1988iqi}
D.~Alde \textit{et al.} [IHEP-Brussels-Los Alamos-Annecy(LAPP)],
``Evidence for a $1^{-+}$Exotic Meson,''
Phys. Lett. B \textbf{205}, 397 (1988).


\bibitem{E852:2001ikk}
E.~I.~Ivanov \textit{et al.} [E852],
``Observation of exotic meson production in the reaction $\pi^- p \rightarrow \eta' \pi^- p$ at 18~GeV/$c$,''
Phys. Rev. Lett. \textbf{86}, 3977-3980 (2001).


\bibitem{E852:2004gpn}
J.~Kuhn \textit{et al.} [E852],
``Exotic meson production in the $f_1(1285)\,\pi^-$ system observed in the reaction $\pi^- p \rightarrow \eta\,\pi^+ \pi^- \pi^- p$ at 18~GeV/$c$,''
Phys. Lett. B \textbf{595}, 109-117 (2004).


\bibitem{BESIII:2022riz}
M.~Ablikim \textit{et al.} [BESIII],
``Observation of an isoscalar resonance with exotic $J^{PC}=1^{-+}$ quantum numbers in $J/\psi \rightarrow \gamma\,\eta\,\eta'$,''
Phys. Rev. Lett. \textbf{129}, no.19, 192002 (2022)
[erratum: Phys. Rev. Lett. \textbf{130}, no.15, 159901 (2023)].



\bibitem{BESIII:2022iwi}
M.~Ablikim \textit{et al.} [BESIII],
``Partial wave analysis of $J/\psi \rightarrow \gamma\,\eta\,\eta'$,''
Phys. Rev. D \textbf{106}, no.7, 072012 (2022)
[erratum: Phys. Rev. D \textbf{107}, no.7, 079901 (2023)].


\bibitem{Chen:2022qpd}
H.~X.~Chen, N.~Su and S.~L.~Zhu,
``QCD axial anomaly enhances the $\eta\,\eta'$ decay of the hybrid candidate $\eta_{1}(1855)$,''
Chin. Phys. Lett. \textbf{39}, no.5, 051201 (2022).


\bibitem{Chen:2023ukh}
B.~Chen, S.~Q.~Luo and X.~Liu,
``Constructing the $J^{PC}=1^{-+}$ light-flavor hybrid nonet with the newly observed $\eta_1(1855)$,''
Phys. Rev. D \textbf{108}, no.5, 054034 (2023).


\bibitem{Chen:2022isv}
F.~Chen, X.~Jiang, Y.~Chen, M.~Gong, Z.~Liu, C.~Shi and W.~Sun,
``$1^{-+}$ hybrid meson in $J/\psi$ radiative decays from lattice QCD,''
Phys. Rev. D \textbf{107}, no.5, 054511 (2023).


\bibitem{Qiu:2022ktc}
L.~Qiu and Q.~Zhao,
``Towards the establishment of the light $J^{P(C)} = 1^{-+}$ hybrid nonet*,''
Chin. Phys. C \textbf{46}, no.5, 051001 (2022).


\bibitem{Shastry:2023ths}
V.~Shastry and F.~Giacosa,
``Radiative production and decays of the exotic $\eta_1'(1855)$ and its siblings,''
Nucl. Phys. A \textbf{1037}, 122683 (2023).


\bibitem{Shastry:2022mhk}
V.~Shastry, C.~S.~Fischer and F.~Giacosa,
``The phenomenology of the exotic hybrid nonet with $\pi_1(1600)$ and $\eta_1(1855)$,''
Phys. Lett. B \textbf{834}, 137478 (2022).


\bibitem{JPAC:2018zyd}
A.~Rodas \textit{et al.} [JPAC],
``Determination of the pole position of the lightest hybrid meson candidate,''
Phys. Rev. Lett. \textbf{122}, no.4, 042002 (2019).


\bibitem{Kopf:2020yoa}
B.~Kopf, M.~Albrecht, H.~Koch, M.~K{\"u}{\ss}ner, J.~Pychy, X.~Qin and U.~Wiedner,
``Investigation of the Lightest Hybrid Meson Candidate with a Coupled-Channel Analysis of $\bar{p}p$-, $\pi^- p$-, and $\pi\pi$-Data,''
Eur. Phys. J. C \textbf{81}, no.12, 1056 (2021).


\bibitem{Dudek:2010wm}
J.~J.~Dudek, R.~G.~Edwards, M.~J.~Peardon, D.~G.~Richards and C.~E.~Thomas,
``Toward the excited meson spectrum of dynamical QCD,''
Phys. Rev. D \textbf{82}, 034508 (2010).


\bibitem{Wan:2022xkx}
B.~D.~Wan, S.~Q.~Zhang and C.~F.~Qiao,
``Possible Structure of the Newly Found Exotic State $\eta_1(1855)$,''
Phys. Rev. D \textbf{106}, no.7, 074003 (2022).



\bibitem{Dong:2022cuw}
X.~K.~Dong, Y.~H.~Lin and B.~S.~Zou,
``Interpretation of the $\eta_1(1855)$ as a $K\bar{K}_1(1400) + \text{c.c.}$ Molecule,''
Sci. China Phys. Mech. Astron. \textbf{65}, no.6, 261011 (2022).



\bibitem{Yan:2023vbh}
M.~J.~Yan, J.~M.~Dias, A.~Guevara, F.~K.~Guo and B.~S.~Zou,
``On the $\eta_1(1855)$, $\pi_1(1400)$ and $\pi_1(1600)$ as Dynamically Generated States and Their SU(3) Partners,''
Universe \textbf{9}, no.2, 109 (2023).


\bibitem{Yang:2022rck}
F.~Yang, H.~Q.~Zhu and Y.~Huang,
``Analysis of the $\eta_1(1855)$ as a $K\bar{K}_1(1400)$ Molecular State,''
Nucl. Phys. A \textbf{1030}, 122571 (2023).

\bibitem{Isgur:1984bm}
N.~Isgur and J.~E.~Paton,
``A Flux Tube Model for Hadrons in QCD,''
Phys. Rev. D \textbf{31}, 2910 (1985).

\bibitem{Isgur:1985vy}
N.~Isgur, R.~Kokoski and J.~Paton,
``Gluonic Excitations of Mesons: Why They Are Missing and Where to Find Them,''
Phys. Rev. Lett. \textbf{54}, 869 (1985).

\bibitem{Close:1994hc}
F.~E.~Close and P.~R.~Page,
``The Production and decay of hybrid mesons by flux tube breaking,''
Nucl. Phys. B \textbf{443}, 233-254 (1995).

\bibitem{Barnes:1995hc}
T.~Barnes, F.~E.~Close and E.~S.~Swanson,
``Hybrid and conventional mesons in the flux tube model: Numerical studies and their phenomenological implications,''
Phys. Rev. D \textbf{52}, 5242-5256 (1995).

\bibitem{Page:1998gz}
P.~R.~Page, E.~S.~Swanson and A.~P.~Szczepaniak,
``Hybrid meson decay phenomenology,''
Phys. Rev. D \textbf{59}, 034016 (1999).

\bibitem{Burns:2006wz}
T.~Burns and F.~E.~Close,
``Hybrid meson properties in Lattice QCD and Flux Tube Models,''
Phys. Rev. D \textbf{74}, 034003 (2006).

\bibitem{Barnes:1982tx}
T.~Barnes, F.~E.~Close and F.~de Viron,
``Q anti-Q G Hermaphrodite Mesons in the MIT Bag Model,''
Nucl. Phys. B \textbf{224}, 241 (1983).

\bibitem{Chanowitz:1982qj}
M.~S.~Chanowitz and S.~R.~Sharpe,
``Hybrids: Mixed States of Quarks and Gluons,''
Nucl. Phys. B \textbf{222}, 211-244 (1983),
[erratum: Nucl. Phys. B \textbf{228}, 588-588 (1983)].

\bibitem{Lacock:1996vy}
P.~Lacock \textit{et al.} [UKQCD],
``Orbitally excited and hybrid mesons from the lattice,''
Phys. Rev. D \textbf{54}, 6997-7009 (1996).

\bibitem{Lacock:1997an}
P.~Lacock \textit{et al.} [UKQCD],
``Hybrid and orbitally excited mesons in quenched QCD,''
Nucl. Phys. B Proc. Suppl. \textbf{63}, 203-205 (1998).

\bibitem{McNeile:1998cp}
C.~McNeile, C.~W.~Bernard, T.~A.~DeGrand, C.~E.~DeTar, S.~A.~Gottlieb, U.~M.~Heller, J.~Hetrick, R.~Sugar and D.~Toussaint,
``Exotic meson spectroscopy from the clover action at beta = 5.85 and beta = 6.15,''
Nucl. Phys. B Proc. Suppl. \textbf{73}, 264-266 (1999).

\bibitem{Lacock:1998be}
P.~Lacock \textit{et al.} [TXL],
``Hybrid and orbitally excited mesons in full QCD,''
Nucl. Phys. B Proc. Suppl. \textbf{73}, 261-263 (1999).

\bibitem{Bernard:2003jd}
C.~Bernard, T.~Burch, E.~B.~Gregory, D.~Toussaint, C.~E.~DeTar, J.~Osborn, S.~A.~Gottlieb, U.~M.~Heller and R.~Sugar,
``Lattice calculation of $1^{-+}  $ hybrid mesons with improved Kogut-Susskind fermions,''
Phys. Rev. D \textbf{68}, 074505 (2003).

\bibitem{Hedditch:2005zf}
J.~N.~Hedditch, W.~Kamleh, B.~G.~Lasscock, D.~B.~Leinweber, A.~G.~Williams and J.~M.~Zanotti,
``$1^{-+}$ exotic meson at light quark masses,''
Phys. Rev. D \textbf{72}, 114507 (2005).

\bibitem{Dudek:2009qf}
J.~J.~Dudek, R.~G.~Edwards, M.~J.~Peardon, D.~G.~Richards and C.~E.~Thomas,
``Highly excited and exotic meson spectrum from dynamical lattice QCD,''
Phys. Rev. Lett. \textbf{103}, 262001 (2009).

\bibitem{Dudek:2011tt}
J.~J.~Dudek, R.~G.~Edwards, B.~Joo, M.~J.~Peardon, D.~G.~Richards and C.~E.~Thomas,
Phys. Rev. D \textbf{83}, 111502 (2011).

\bibitem{Dudek:2011bn}
J.~J.~Dudek,
``The lightest hybrid meson supermultiplet in QCD,''
Phys. Rev. D \textbf{84}, 074023 (2011).

\bibitem{Dudek:2013yja}
J.~J.~Dudek \textit{et al.} [Hadron Spectrum],
``Toward the excited isoscalar meson spectrum from lattice QCD,''
Phys. Rev. D \textbf{88}, no.9, 094505 (2013).

\bibitem{Balitsky:1982ps}
I.~I.~Balitsky, D.~Diakonov and A.~V.~Yung,
``EXOTIC MESONS WITH $J^{\mathrm{PC}} = 1^{-+}$ FROM QCD SUM RULES,''
Phys. Lett. B \textbf{112}, 71-75 (1982).

\bibitem{Govaerts:1983ka}
J.~Govaerts, F.~de Viron, D.~Gusbin and J.~Weyers,
``Exotic mesons from QCD sum rules,''
Phys. Lett. B \textbf{128}, 262 (1983),
[erratum: Phys. Lett. B \textbf{136}, 445-445 (1984)].

\bibitem{Govaerts:1984hc}
J.~Govaerts, L.~J.~Reinders, H.~R.~Rubinstein and J.~Weyers,
``Hybrid quarkonia from QCD sum rules,''
Nucl. Phys. B \textbf{258}, 215-229 (1985).

\bibitem{Govaerts:1985fx}
J.~Govaerts, L.~J.~Reinders and J.~Weyers,
``Radial Excitations and Exotic Mesons via {QCD} Sum Rules,''
Nucl. Phys. B \textbf{262}, 575-592 (1985).

\bibitem{Govaerts:1986pp}
J.~Govaerts, L.~J.~Reinders, P.~Francken, X.~Gonze and J.~Weyers,
``Coupled {QCD} Sum Rules for Hybrid Mesons,''
Nucl. Phys. B \textbf{284}, 674 (1987).

\bibitem{Zhu:1998ki}
S.~L.~Zhu,
``Hybrid quarkonium masses up to the order of O(1 / m(Q)),''
Phys. Rev. D \textbf{60}, 031501 (1999).

\bibitem{Jin:2002rw}
H.~Y.~Jin, J.~G.~Korner and T.~G.~Steele,
``Improved determination of the mass of the $1^{-+}$ light hybrid meson from QCD sum rules,''
Phys. Rev. D \textbf{67}, 014025 (2003).

\bibitem{Qiao:2010zh}
C.~F.~Qiao, L.~Tang, G.~Hao and X.~Q.~Li,
``Determining $1^{--}$ Heavy Hybrid Masses via QCD Sum Rules,''
J. Phys. G \textbf{39}, 015005 (2012).

\bibitem{Berg:2012gd}
R.~Berg, D.~Harnett, R.~T.~Kleiv and T.~G.~Steele,
``Mass Predictions for Pseudoscalar $J^{PC}=0^{-+}$ Charmonium and Bottomonium Hybrids in QCD Sum-Rules,''
Phys. Rev. D \textbf{86}, 034002 (2012).

\bibitem{Chen:2013zia}
W.~Chen, R.~T.~Kleiv, T.~G.~Steele, B.~Bulthuis, D.~Harnett, J.~Ho, T.~Richards and S.~L.~Zhu,
``Mass Spectrum of Heavy Quarkonium Hybrids,''
JHEP \textbf{09}, 019 (2013).

\bibitem{Chen:2013pya}
W.~Chen, H.~y.~Jin, R.~T.~Kleiv, T.~G.~Steele, M.~Wang and Q.~Xu,
``QCD sum-rule interpretation of X(3872) with $J^{PC}=1^{++}$ mixtures of hybrid charmonium and $\overline{D}D^*$ molecular currents,''
Phys. Rev. D \textbf{88}, no.4, 045027 (2013).

\bibitem{Chen:2013eha}
W.~Chen, T.~G.~Steele and S.~L.~Zhu,
``Masses of the bottom-charm hybrid $\bar bGc$ states,''
J. Phys. G \textbf{41}, 025003 (2014).

\bibitem{Li:2021fwk}
S.~H.~Li, Z.~S.~Chen, H.~Y.~Jin and W.~Chen,
``Mass of $1^{-+}$ four-quark-hybrid mixed states,''
Phys. Rev. D \textbf{105}, no.5, 054030 (2022).

\bibitem{Palameta:2017ols}
A.~Palameta, J.~Ho, D.~Harnett and T.~G.~Steele,
``QCD sum-rules analysis of vector ($1^{--}$) heavy quarkonium meson-hybrid mixing,''
Phys. Rev. D \textbf{97}, no.3, 034001 (2018).

\bibitem{Palameta:2018yce}
A.~Palameta, D.~Harnett and T.~G.~Steele,
``Meson-Hybrid Mixing in $J^{PC}=1^{++}$ Heavy Quarkonium from QCD Sum-Rules,''
Phys. Rev. D \textbf{98}, no.7, 074014 (2018).

\bibitem{Ho:2018cat}
J.~Ho, R.~Berg, W.~Chen, D.~Harnett and T.~G.~Steele,
``Mass Calculations of Light Quarkonium, Exotic $J^{PC}=0^{+-}$ Hybrid Mesons from Gaussian Sum-Rules,''
Phys. Rev. D \textbf{98}, no.9, 096020 (2018).

\bibitem{Ho:2019org}
J.~Ho, R.~Berg, T.~G.~Steele, W.~Chen and D.~Harnett,
``Is the $Y(2175)$ a Strangeonium Hybrid Meson?,''
Phys. Rev. D \textbf{100}, no.3, 034012 (2019).

\bibitem{Barsbay:2024vjt}
B.~Barsbay, K.~Azizi and H.~Sundu,
``Light quarkonium hybrid mesons,''
Phys. Rev. D \textbf{109}, no.9, 094034 (2024).

\bibitem{Esmer:2025xss}
G.~D.~Esmer, K.~Azizi, H.~Sundu and S.~T{\"u}rkmen,
``Decays of the light hybrid meson $1^{-+}$,''
Phys. Rev. D \textbf{111}, no.3, 034041 (2025).

\bibitem{Agaev:2025llz}
S.~S.~Agaev, K.~Azizi and H.~Sundu,
``Tensor hybrid charmonia,''
Phys. Rev. D \textbf{112}, no.1, 014003 (2025).

\bibitem{Agaev:2025mqe}
S.~S.~Agaev, K.~Azizi and H.~Sundu,
``Tensor hybrid mesons $\bar{b}gc$,''
Phys. Lett. B \textbf{868}, 139732 (2025).

\bibitem{HadronSpectrum:2012gic}
L.~Liu \textit{et al.} [Hadron Spectrum],
``Excited and exotic charmonium spectroscopy from lattice QCD,''
JHEP \textbf{07}, 126 (2012).

\bibitem{Meyer:2015eta}
C.~A.~Meyer and E.~S.~Swanson,
``Hybrid Mesons,''
Prog. Part. Nucl. Phys. \textbf{82}, 21-58 (2015).

\bibitem{Wang:2023whb}
Q.~N.~Wang, D.~K.~Lian and W.~Chen,
``Predictions of the hybrid mesons with exotic quantum numbers $J^{PC}=2^{+-}$,''
Phys. Rev. D \textbf{108}, no.11, 114010 (2023).

\bibitem{Su:2022fqr}
N.~Su, H.~X.~Chen, W.~Chen and S.~L.~Zhu,
``Light double-gluon hybrid states from QCD sum rules,''
Phys. Rev. D \textbf{107}, no.3, 034010 (2023).

\bibitem{Su:2023jxb}
N.~Su, W.~H.~Tan, H.~X.~Chen, W.~Chen and S.~L.~Zhu,
``Light double-gluon hybrid states with the exotic quantum numbers $J^{PC}=1^{-+}$ and $ 3^{-+} $,''
Phys. Rev. D \textbf{107}, no.11, 114005 (2023).

\bibitem{Lian:2024fsb}
D.~K.~Lian, Q.~N.~Wang, X.~L.~Chen, P.~F.~Yang, W.~Chen and H.~X.~Chen,
``Revisit the heavy quarkonium double-gluon hybrid mesons with exotic quantum numbers,''
JHEP \textbf{06}, 173 (2024).

\bibitem{Chen:2021smz}
H.~X.~Chen, W.~Chen and S.~L.~Zhu,
``New hadron configuration: The double-gluon hybrid state,''
Phys. Rev. D \textbf{105}, no.5, L051501 (2022).

\end{thebibliography}
\end{document}